\documentclass[preprint,aps]{revtex4}
\usepackage{graphicx}
\usepackage[dvips]{epsfig}
\newcounter{fig}

\begin{document}
\title{Damping of coupled phonon--plasmon modes}

%%%%%%

\author{L. A. Falkovsky}
\email{falk@itp.ac.ru}
\affiliation{Landau Institute for Theoretical Physics, 117337 Moscow, Russia}

%\email{falk@itp.ac.ru}

\bigskip
\bigskip
%\maketitle

\begin{abstract}
The effect of free carriers on  dispersion and  damping of coupled
phonon-plasmon modes  is considered in the long-wave
approximation. The electron and
phonon scattering rate as well as  Landau damping are taken into account.
\end{abstract}

\maketitle

\newcounter{newcnt}
\section{Introduction}

After the pioneering paper by Migdal \cite{Mi},
the effect of  electron-phonon interactions on the phonon dispersion
has been permanently debated.
The problem was how to explain that  the many-body
approach based on the Fr\"{o}hlich Hamiltonian
gives a strong phonon renormalization. For instance,  the
 sound velocity  is renormalized
on the order of  phonon-electron coupling
$\lambda$ which has the  value of unity.
It follows that the phonon--electron system can be  unstable.
The phonon softening caused by the electron-phonon
interaction  have been discussed in many papers.
All these results contradict
to the Born-Oppenheimer conception \cite{BHK}
according to which the light electrons
must follow the slow lattice vibrations. Therefore, the phonon
renormalization  should involve the small parameter of nonadiabacity
 $\sqrt{m/M}$, where $m$ and $M$ are the electron and ion masses
correspondingly.

The discrepancy was resolved by Brovman and Kagan \cite{BK}.
They demonstrated the shortcomings of the Fr\"ohlich model.
Employing the adiabatic approximation,
they found that there are two terms in the second order
perturbation theory, which compensate each other
making a result small by the nonadiabatic parameter.

But recently, Alexandrov and Schrieffer \cite{AS}
again obtained the strong phonon renormalization and
predicted an extremely large dispersion of optical phonons
(on  order of the Fermi velocity)
because of coupling to electrons. The large phonon dispersion is
a typical result of the theory \cite{AGD} used  the Fr\"olich
Hamiltonian. Reizer \cite{R}  stressed the
importance of screening effects accompanying the longitudinal
optical modes, but the nonphysical renormalization remained
in his results. For the first time, the screening
of the Coulomb field in optical vibrations was treated in the paper
\cite{GLF}.

Only the phonon frequencies were  calculated
in the works \cite{AS}, \cite{R}, \cite{GLF},
with no results available for the attenuation of optical phonons.
The electron collisions  with each other, with defects
and phonons were ignored. Beside the electron collision rate $\gamma$,
 the natural phonon width $\Gamma^{\text{nat}}$ was also disregarded.
The natural phonon width is caused by  the anharmonicity processes
of the phonon decay into two (or more) phonons.
Notice, that the collision processes determine the conductivity and
the dielectric permittivity, i. e. electrodynamics of the electron-phonon
system. In the optic range, the collision rates
$\gamma, \Gamma^{\text{nat}} \sim \sqrt{m/M} \omega_{\text{O}}$
are small  compared with a typical phonon frequency $\omega_{\text{O}}$
and they give the widths of the plasmon and phonon resonances. Experimental
studies of this resonances provide informations about  isotopic
compositions and  quality of semiconductor materials.

Using the Boltzmann equation for electrons and the equation of motion
for phonons we  calculated
\cite{Fa1} the frequency shift and the width of optical phonons in metals.
We  take into account all the mentioned features:
the electron--phonon interaction  of different forms,
 the Coulomb screening, and the collision rates
of electrons and phonons.  We find that
 dispersion and  damping  of the longitudinal modes can be correctly
described
if we neglect the direct electron-phonon interaction
$\lambda$ and keep only the  screening and the collision rates
  $\gamma$, $\Gamma^{\text{nat}}$.
Emphasize, that in the semiclassical approximation, when the phonon momentum
transferred to electrons is small compared
with the electron momentum,  the method of the Boltzmann equation is
completely equivalent to the diagram technique. The corresponding equations
may be formulated as  equations for the electron and phonon self-energies.
But in any case, we must properly  incorporate
adiabatic approximation and  screening.

Experimentally, it is  convenient to investigate the free-carrier effect
on phonon modes, varying the carrier concentration, i. e. for doped
semiconductors or superconductors (see, for example,  the recent work
\cite{AMG},  carried out on the HTcS compound
Nd$_{1.86}$Ce$_{0.14}$CuO$_{4+\delta}$ using IXS).
Therefore, in the present paper,
we extend the results of the paper \cite{Fa1} to the
case of  small free-carrier concentrations when the electron plasma
frequency $\omega_{ep}$ can be of the order of
the phonon frequency $\omega_{O}$,
and the coupled phonon-plasmon modes can exist.
We consider  the screening as well as the
collision effects $\gamma$ and $\Gamma^{\text{nat}}$,
focusing on the width of the phonon-plasmon modes.
We simplify essentially the problem ignoring
the direct electron-phonon interaction
(terms with $\lambda$) and
 assuming that, first, the electron system is degenerate and, second,
the momentum
transfer $k$ is small compared with the Fermi momentum $p_F$.

\section{Asymptotic expressions for the dielectric function}

Let us find the limiting expressions of the dielectric function
\begin{equation}\label{dfu}
\varepsilon(k,\omega)=\varepsilon_{\infty}\frac
{\omega^2-\omega_{\text{LO}}^2+i\Gamma^{\text{nat}}}
{\omega^2-\omega_{\text{TO}}^2+i\Gamma^{\text{nat}}}
-\frac{4\pi e^2\nu_0
\langle v_z/
\Delta_p( k)\rangle}{k(1-i\langle \gamma/\Delta_p( k)\rangle)}.
\end{equation}
Here the ion contribution (first term)  is assumed as dispersionless,
while we are interested in  small wave vectors in comparison with
the Brillouin-zone size. This term has the pole and the  zero at the
frequencies of the transverse and longitudinal phonons correspondingly.
The natural phonon width $i\Gamma^{\text{nat}}/2$ is added to $\omega$;
the high-frequency ion permittivity is denoted $\varepsilon_{\infty}$.

The second term in Eq. (\ref{dfu})
is the electron contribution into permittivity
for $k\ll p_F$, written  \cite{Fa1} with the help of
the Boltzmann equation in  approximation
of the collision rate $\gamma$; the notation
$$\Delta_p( k)=\omega-{\bf k}{\bf v}+i\gamma$$
is introduced.
The angular brackets
\begin{displaymath}
\langle...\rangle = \frac{1}{\nu_0}\int(...){2dS_{F}\over v(2\pi)^3}.
\end{displaymath}
denote  averaging over the Fermi surface with the density of states $\nu_0$.
For the isotropic case, the density of states on the Fermi surface
 $\nu_0 = m^* p_F/\pi^2$ and $m^*$ is the effective
electron mass.

Let us rewrite the electron contribution
\begin{equation}\label{dfu1}
\varepsilon_e(k,\omega)=\varepsilon_{\infty}\frac{k_0^2}{k^2 }
\left[1-\frac{\langle \omega/\Delta_p(k)\rangle}
{1-i\langle \gamma/\Delta_p(k)\rangle}\right],
\end{equation}
where the Thomas-Fermi parameter
$k_0^2=4\pi e^2\nu_0/\varepsilon_{\infty}$.

For the isotropic Fermi surface we can carry out the integration
$$\langle 1/\Delta_p(k)\rangle=
\frac{1}{2 kv_F}\ln{\frac{\omega+i\gamma+kv_F}
{\omega+i\gamma-kv_F}}.$$
Separating the imaginary and real parts we have
\begin{equation} \label{del}
\langle 1/\Delta_p(k)\rangle=
\frac{1}{2kv_F}\left[\frac{1}{2}\ln\frac
{(\omega+kv_F)^2+\gamma^2}
{(\omega-kv_F)^2+\gamma^2}+i
\arctan\frac{\omega-kv_F}{\gamma}-i
\arctan\frac{\omega+kv_F}{\gamma}\right].
\end{equation}

The imaginary part
 known as the  Landau damping
is  pronounced at
$kv_F>|\omega+i\gamma|$:
\begin{equation} \label{epsk}
\varepsilon_e(k,\omega)= \varepsilon_{\infty}
(k_0/ k)^2(1+i\pi\omega/2 kv_F),  \quad kv_F>>|\omega+i\gamma|.
\end{equation}

In the range $\gamma<\omega-kv_F<<kv_F$, we have
\begin{equation} \label{eps1}
\varepsilon_e(k,\omega)= \varepsilon_{\infty}\frac{k_0^2}{k^2}\left\{
1-\frac{\omega}{2 kv_F}\left[\frac{1}{2}\ln{\frac{4k^2v_F^2}
{(\omega-kv_F)^2+\gamma^2}}-i\frac{\gamma}{\omega-kv_F}\right]\right\}.
\end{equation}

For the small
$ kv_F<<|\omega+i\gamma|$,  the expansion in
$ kv_F/(\omega+i\gamma)$
gives
\begin{equation} \label{epso}
\varepsilon_e(k,\omega)= \varepsilon_{\infty}
\left\{1
-{\omega_{pe}^2\over \omega(\omega+i\gamma)}\left[1+\frac{3}{5}
\left(\frac{kv_F}{\omega+i\gamma}\right)^2\right]\right\}.
\end{equation}
Here the $k$-independent term represents the Drude conductivity.
The electron plasma frequency  is given by the integral over the Fermi
surface
$$\omega_{pe}^2=\frac{e^2}{3\pi^2
\varepsilon_{\infty}}\int vdS_F.$$

The limiting expressions (\ref{epsk})--(\ref{epso}) are also valid
for the arbitrary Fermi surface, but the constant $v_F$ has different
values.
In Eq. (\ref{epsk}), this is
 an average velocity on the belt ${\bf v}\perp {\bf k}$,
 the velocity in the limiting point
${\bf v}\parallel {\bf k}$
in Eq. (\ref{eps1}) , and
the squared velocity  averaged  over the whole Fermi surface
in Eq. (\ref{epso}).
We do not pay attention the cases when the Fermi surface have
flat or cylindrical pieces.

\section{Frequency and damping of  phonon--plasmon modes}

The frequencies of the longitudinal phonon--plasmon modes
are determined by the equation
$\varepsilon(k,\omega)=0$.
In the absence of the electron and phonon collisions
($\gamma=\Gamma^{\text{nat}}=0$),
we obtain with the help of Eqs. (\ref{dfu})  and (\ref{epso})
the biquadratic equation. It gives
 the frequencies of the coupled phonon-plasmon modes at
$k=0$:
\begin{equation} \label{pm}
\omega^2_{\pm}=
\frac{1}{2}
(\omega_{pe}^2+\omega_{\text{LO}}^2)
\pm\frac{1}{2}\left[
(\omega_{pe}^2+\omega_{\text{LO}}^2)^2-4\omega_{pe}^2\omega_{\text{TO}}^2
\right]^{1/2}.
\end{equation}
These frequencies (related to the $\omega_{\text{TO}}$) are shown
in Fig. 1 (left panel) as functions of the electron concentration,
 namely, $\omega_{pe}/\omega_{\text{TO}}$.
The upper dashed line begins at
$\omega_{\text{LO}}$ and tends to the electron plasma frequency
$\omega_{pe}$.
The lower frequency (solid line) starts as
$\omega_{pe}\omega_{\text{TO}}/\omega_{\text{LO}}$ and then approaches
 $\omega_{\text{TO}}$.
In other words,
observing in the optic range the longitudinal phonon mode and
adding electrons,
we see a transition of the longitudinal phonon frequency
from
$\omega_{\text{LO}}$  to $\omega_{\text{TO}}$. This is a result of
the Coulomb screening.

Since  both the collision rates $\gamma$ and $\Gamma^{\text{nat}}$
in the optic range
are small in comparison with
$\omega_{\text{O}}$, the damping of the phonon-plasmon modes can be
added to their frequencies
$\omega=\omega_{\pm}-i\Gamma_{\pm}/2$.
Using Eqs. (\ref{dfu}) and (\ref{pm}) we find at $k=0$:
$$\Gamma_{+}=[\gamma(\omega_{+}^2-\omega^2_{\text{LO}})+
\Gamma^{\text{nat}}(\omega_{+}^2-\omega_{pe}^2)]/
(\omega_{+}^2-\omega_{-}^2),$$
$$
\Gamma_{-}=[\gamma(\omega_{-}^2-\omega^2_{\text{LO}})+
\Gamma^{\text{nat}}(\omega_{-}^2-\omega_{pe}^2)]/
(\omega_{-}^2-\omega_{+}^2).$$
The behavior of  damping as a function of the electron concentration
is shown in Fig. 1 (right panel)  for the  ratio
$\gamma/\Gamma^{\text{nat}}=3$.
We see that the concentration dependence of  widths
corresponds with the type of modes: for the low electron concentration,
the solid line represents mainly the electron plasma mode and the
dashed line    is associated with the longitudinal phonon. The character
of modes reverses at the higher electron concentrations.

Now let us consider the dispersion of the coupled phonon--plasmon
modes. When the wave vector $k$ increases,
the upper mode (see Fig. 2 and Eq.
(\ref{eps1}))
approaches the asymptote
$\omega=kv_F$:
$$\omega_{+}=kv_F\left\{1+2\exp\left[-2-\frac{k^2(k^2v_F^2-\omega^2_{\text{LO}})}
{k_0^2(k^2v_F^2-\omega^2_{\text{TO}})}\right]\right\}-i\gamma.$$
The lower mode drops into the domain
$kv_F>\omega$ (dashed curve),
where the Landau damping appears:
\begin{equation}\label{ld}
\omega_{-}^2=\frac{k^2\omega_{\text{LO}}^2+k_{0}^2\omega_{\text{TO}}^2}
{k^2+k_0^2} - i\frac{\pi \omega k k_0^2\omega_{pi}^2}{2v_F(k^2+k_0^2)^2}.
\end{equation}
Here the ion plasma frequency $\omega_{pi}^2=\omega^2_{\text{LO}}-
\omega^2_{\text{TO}}$.
In the imaginary part,
we should substitute instead of $\omega$
the frequency $\omega_{-}$ defined by the real part.
Let us emphasize, that the large dispersion arises under the conditions
$k_0v_F\gg\omega$ and $k_0<p_F$. Then,
at $k\simeq k_0$,
the frequency changes from
$\omega_{\text{TO}}$ to $\omega_{\text{LO}}$.
Again this is the effect of  screening.

In the general case, the frequency and the damping of
 phonon--plasmon modes  can be found in the numerical solution of the
equation
$\varepsilon(k,\omega)=0$.
But of particular interest is the function
$-\text{Im}~1/\varepsilon(k,\omega)$,
since it gives the intensity of the inelastic Raman or X-ray
scattering, where $\omega$ and ${\bf k}$ have the sense
of frequency and momentum transfers, respectively.
The plots of the intensity, obtained with the help of
Eqs. (\ref{dfu}) -- (\ref{del}) are shown
in Figs. 3 -- 5 for the cases of the large electron density
($\omega_{pe}>\omega_{\text{TO}}$). The peak at
$\omega/\omega_{\text{TO}}=2.6\div 2.7$
corresponds to the plasmon excitation; notice, that
 $\omega_{pe}=k_0v_F/\sqrt{3}$ for the quadratic electron dispersion.

In the domain $\omega/\omega_{\text{TO}}<1$,
there is a phonon peak (see left panel of Fig. 1 and Eq.
(\ref{pm})). Because of the screening, its frequency is smaller than
$\omega_{\text{LO}}$.
 For the small $k$ (Fig. 3), the intensity of the phonon peak is 50 times
less than the plasmon-peak intensity. For the larger $k$ (see Fig. 4),
the broad continuum $\omega<kv_F$
appears on the low-frequency side of the phonon peak. Here the dielectric
 constant $\varepsilon(k,\omega)$
has a noticeable imaginary part (\ref{epsk})
arising from the electron-hole excitations. The intensity of the phonon peak
decreases, its line shape becomes asymmetric similar to the Fano resonance.
Finally, the phonon peak is spread, being immersed in the electron
continuum (see Fig. 5 and Eq. (\ref{ld})).

For the relative low electron density,
$\omega_{pe}<\omega_{\text{TO}}$, the intensities
are shown in Figs. 6 -- 8.   Now the phonon peak is approximately at
$\omega/\omega_{\text{TO}}=1.5$. That yields
 $\omega=\omega_{\text{LO}}$, since we set
$\omega_{\text{LO}}/\omega_{\text{TO}}=\sqrt{2}$.
There are also the plasmon peak and the electron-hole continuum.
In Fig. 6 the plasmon peak is  broader than the phonon peak, since
we put
$\gamma/\Gamma^{\text{nat}}=3$.
 The plasmon peak becomes more broader in Fig. 7
 because of the neighboring electron continuum at
 $\omega<kv_F$. It disappears completely in Fig. 8.

\section{Conclusion}

We have shown that the width of the longitudinal phonon-plasmon modes
increases with increasing of
the free-carrier concentration because their frequency
approaches  the region $kv_F>\omega$, where the electron-hole excitations
exist. The detailed experimental studies could help to clarify the
comparative role of the different mechanisms of the phonon shift and damping
in the metallic and doped semiconducting materials under the influence of
the electron concentration.

The work was partially supported by the RFBR
(project 01-02-16211).

\newpage
{\bf Figure captions}: \\[0.3cm]
\\Fig. 1. Frequencies (in units of $\omega_{\text{TO}}$,  left panel)
and widths (in units of $\Gamma^{\text{nat}}$,  right panel)
of the phonon-plasmon modes at $k=0$
in dependence of the free carrier concentration, namely, of
 the electron plasma frequency
(in units of $\omega_{\text{TO}}$). We set the ratio of the LO and TO
frequencies in the absence of the free carriers
$\omega_{\text{LO}}/\omega_{\text{TO}}=\sqrt{2}$ and the ratio
of the electron and phonon damping  $\gamma/\Gamma^{\text{nat}}=3$.
\\Fig. 2.
The dispersion of phonon-plasmon modes for the metallic
($\omega_{pe}>\omega_{\text{TO}}$, left panel) and
semiconducting ($\omega_{pe}<\omega_{\text{TO}}$, right panel)
 carrier concentration.
The dashed straight lines separate the domain
$kv_F>\omega$, where the Landau damping exists;
the dashed curves represent the strongly damped modes.
\\Fig. 3 -- 8.
The inelastic light scattering intensity
$-\text{Im}~1/\varepsilon(k,\omega)$
as a function of the frequency transfer $\omega$ for
the momentum transfer $k$ and the Thomas-Fermi
parameter $k_0$ (in the units of $\omega_{\text{TO}}/v_F$) shown in the
figures. We set
$\omega_{\text{LO}}/\omega_{\text{TO}}=\sqrt{2}$,
$\Gamma^{\text{nat}}/\omega_{\text{TO}}=10^{-2}$,
$\gamma/\Gamma^{\text{nat}}=3$.
\bigskip

\clearpage
\begin{figure}
\begin{center}
\epsfxsize=150mm
\epsfysize=100mm
\epsfbox{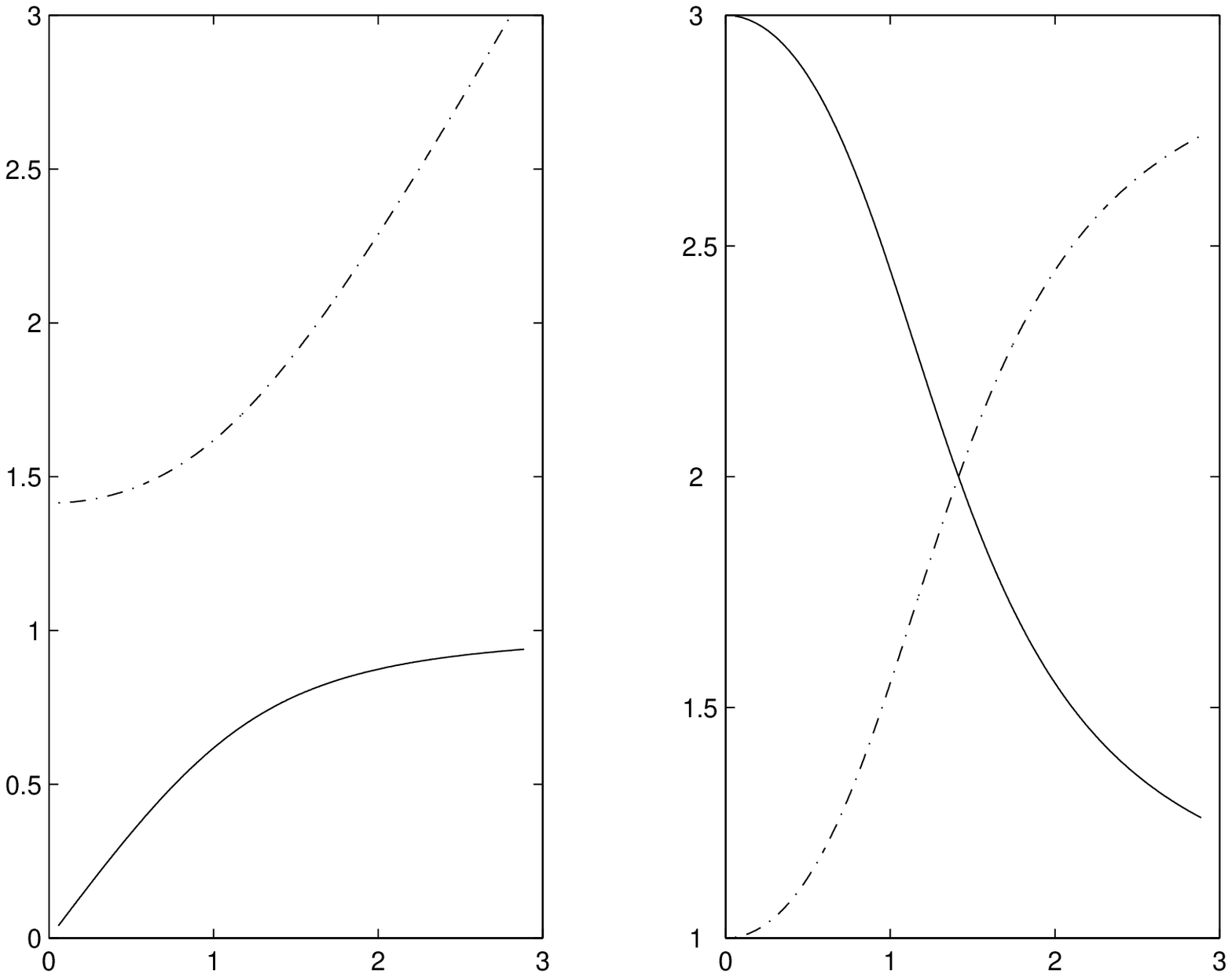}
\begin{picture}(0,-00)
\setlength{\unitlength}{1.000pt}
\put(-110,240){\Large $\omega_{+}$}
\put(-90,80){\Large $\omega_{-}$}
\put(-225,120){\rotatebox{90}{frequency, rel. units}}
\put(-130,-5){\Large $\omega_{pe}$ ,}
\put(-90,-5){rel. units}
\put(160,180){\Large $\Gamma_{+}$}
\put(160,110){\Large $\Gamma_{-}$}
\put(10,120){\rotatebox{90}{width, rel. units}}
\put(100,-5){\Large $\omega_{pe}$ ,}
\put(140,-5){rel. units}

\end{picture}
\end{center}
\vspace{0mm}
\caption{}
\end{figure}

\clearpage
\begin{figure}
\begin{center}
 \epsfxsize=120mm
 \epsfysize=70mm
 \epsfbox{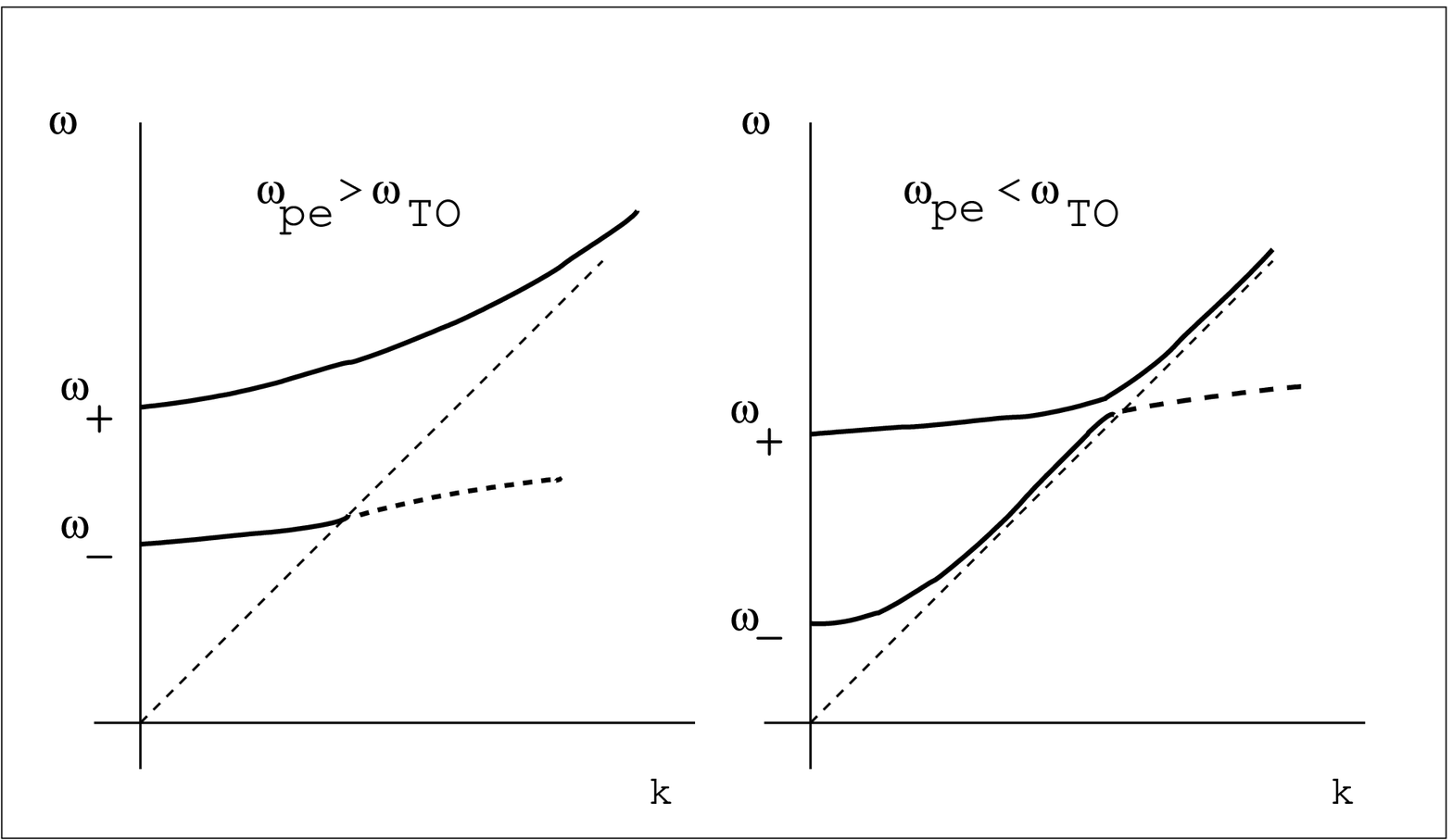}
 \end{center}
 \caption{ }
 \end{figure}

\clearpage
\begin{figure}
\begin{center}
\epsfxsize=140mm
\epsfysize=110mm
\epsfbox{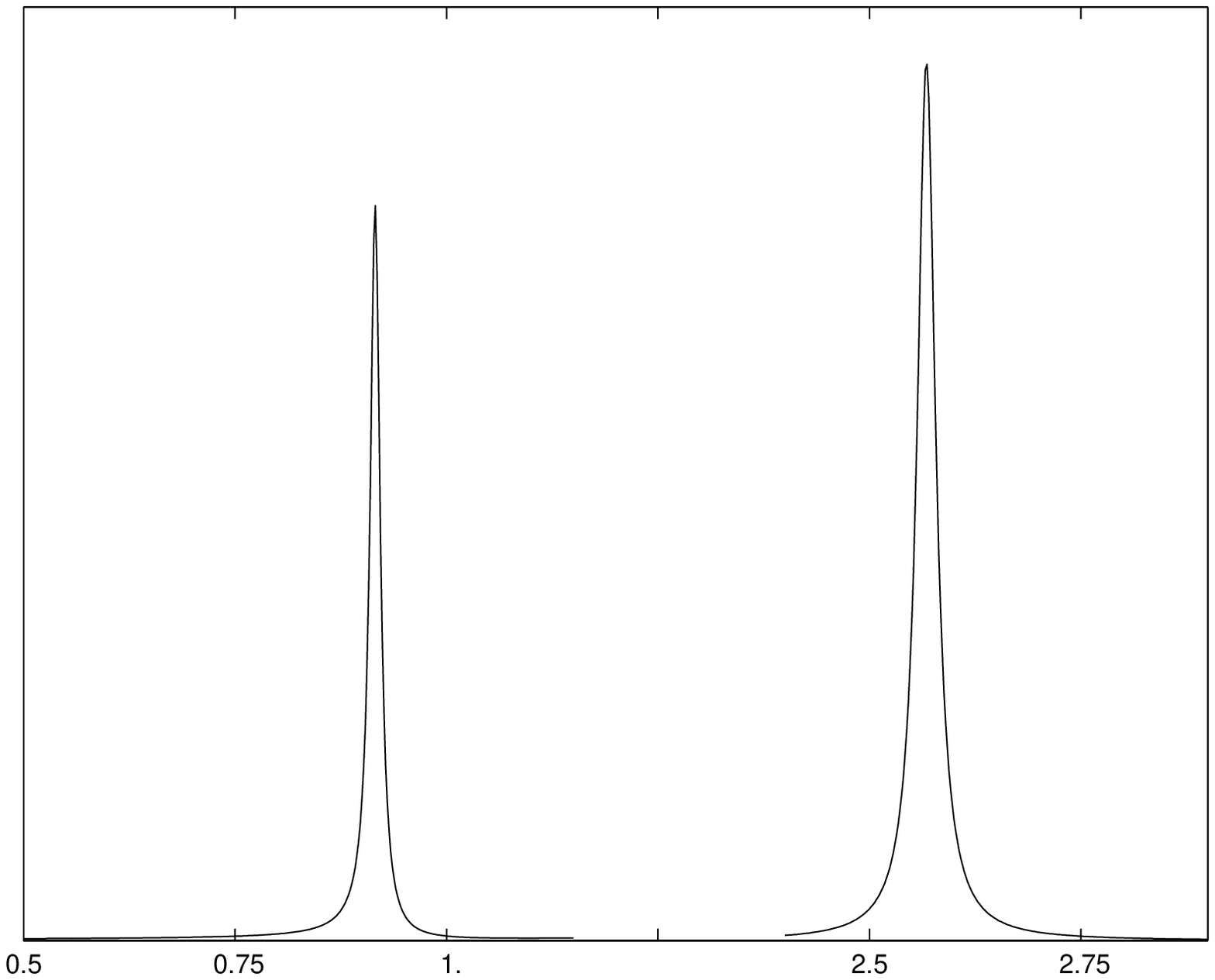}
\begin{picture}(0,-00)
\setlength{\unitlength}{1.000pt}
\put(-150,280){\Large $\omega_{+}$}
\put(-240,190){\Large $\omega_{-}$}
\put(-360,250){\Large $k_{0}=4$}
\put(-360,220){\Large $k=0.4$}
\put(-260,120){\Large $\leftarrow$ }
\put(-240,120){\Large $\times  50$}
\put(-415,100){\rotatebox{90}{intensity}}
\put(-260,-20){frequency transfer, rel. units}
\end{picture}
\end{center}
\caption{ }
\end{figure}

\clearpage
\begin{figure}
\begin{center}
\epsfxsize=140mm
\epsfysize=110mm
\epsfbox{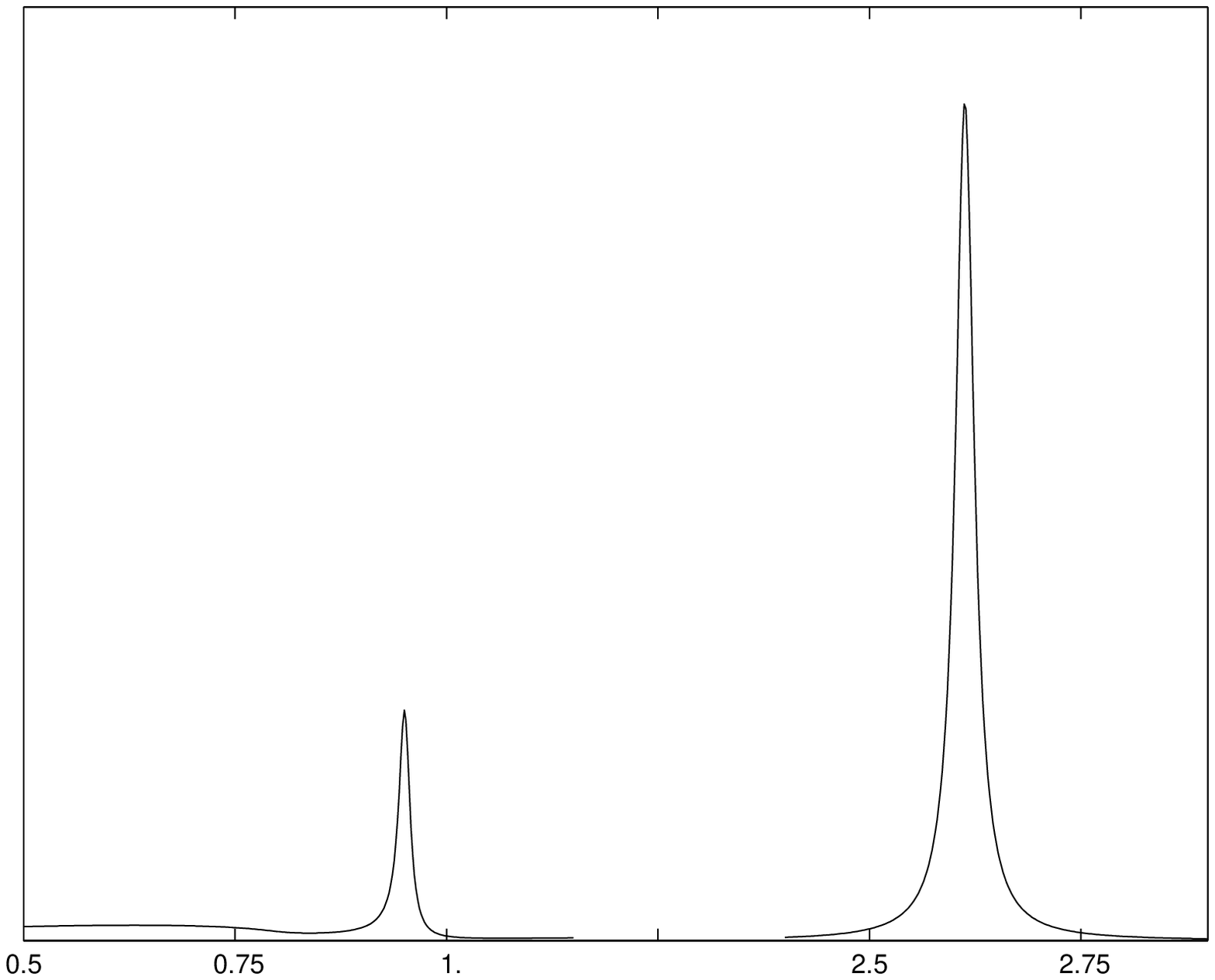}
\begin{picture}(0,-00)
\setlength{\unitlength}{1.000pt}
\put(-150,260){\Large $\omega_{+}$}
\put(-240,100){\Large $\omega_{-}$}
\put(-360,250){\Large $k_{0}=4$}
\put(-360,220){\Large $k=0.8$}
\put(-240,50){\Large $\leftarrow$ }
\put(-220,50){\Large $\times  50$}
\put(-415,100){\rotatebox{90}{intensity}}
\put(-260,-20){frequency transfer, rel. units}
\end{picture}
\end{center}
\caption{ }
\end{figure}

\clearpage
\begin{figure}
\begin{center}
\epsfxsize=140mm
\epsfysize=110mm
\epsfbox{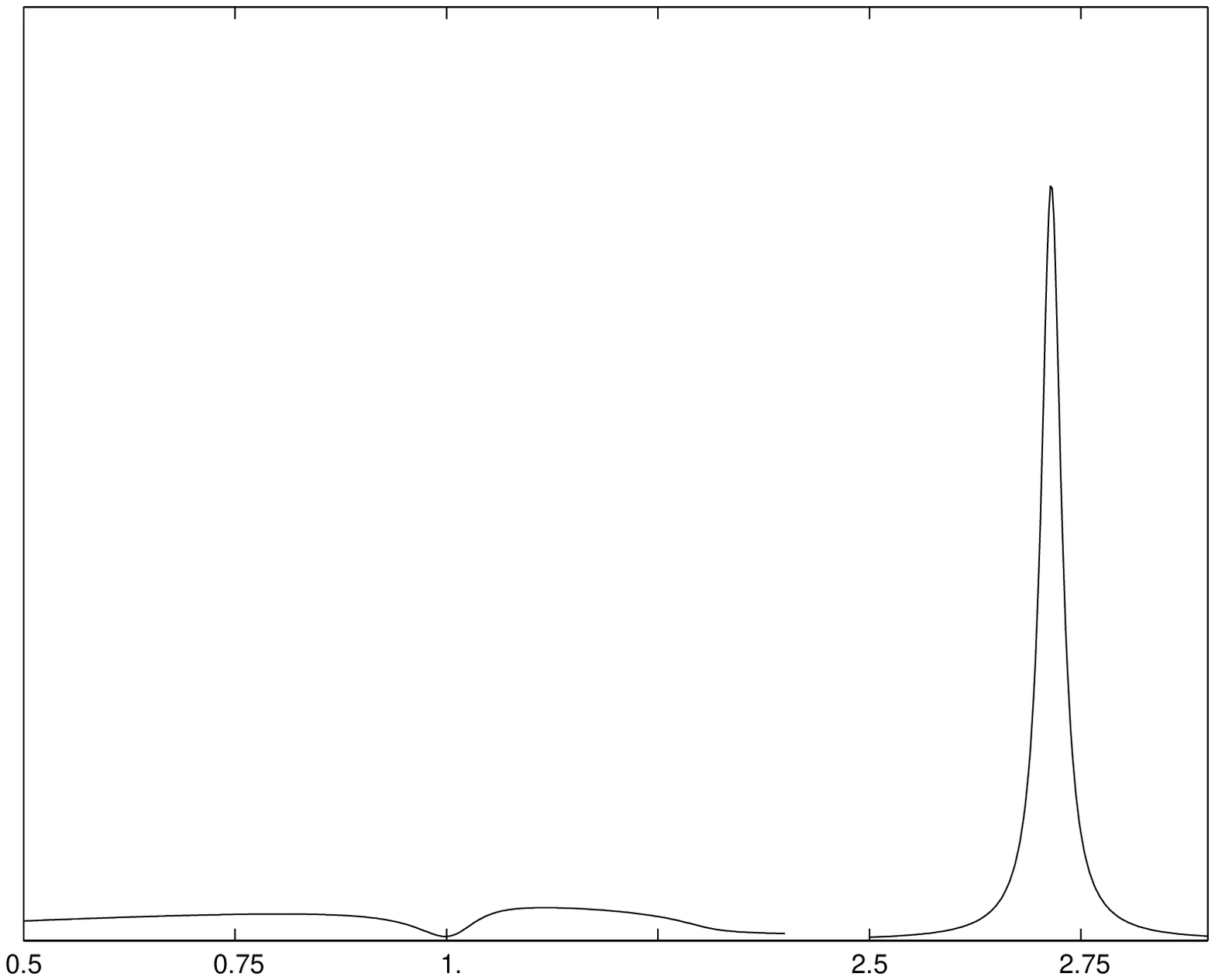}
\begin{picture}(0,-00)
\setlength{\unitlength}{1.000pt}
\put(-110,260){\Large $\omega_{+}$}
\put(-230,80){\Large $\omega_{-}$}
\put(-360,250){\Large $k_{0}=4$}
\put(-360,220){\Large $k=1.3$}
\put(-230,60){\Large $\downarrow$ }
\put(-210,60){\Large $\times  50$}
\put(-415,100){\rotatebox{90}{intensity}}
\put(-260,-20){frequency transfer, rel. units}
\end{picture}
\end{center}
\caption{ }
\end{figure}

\clearpage
\begin{figure}
\begin{center}
\epsfxsize=140mm
\epsfysize=110mm
\epsfbox{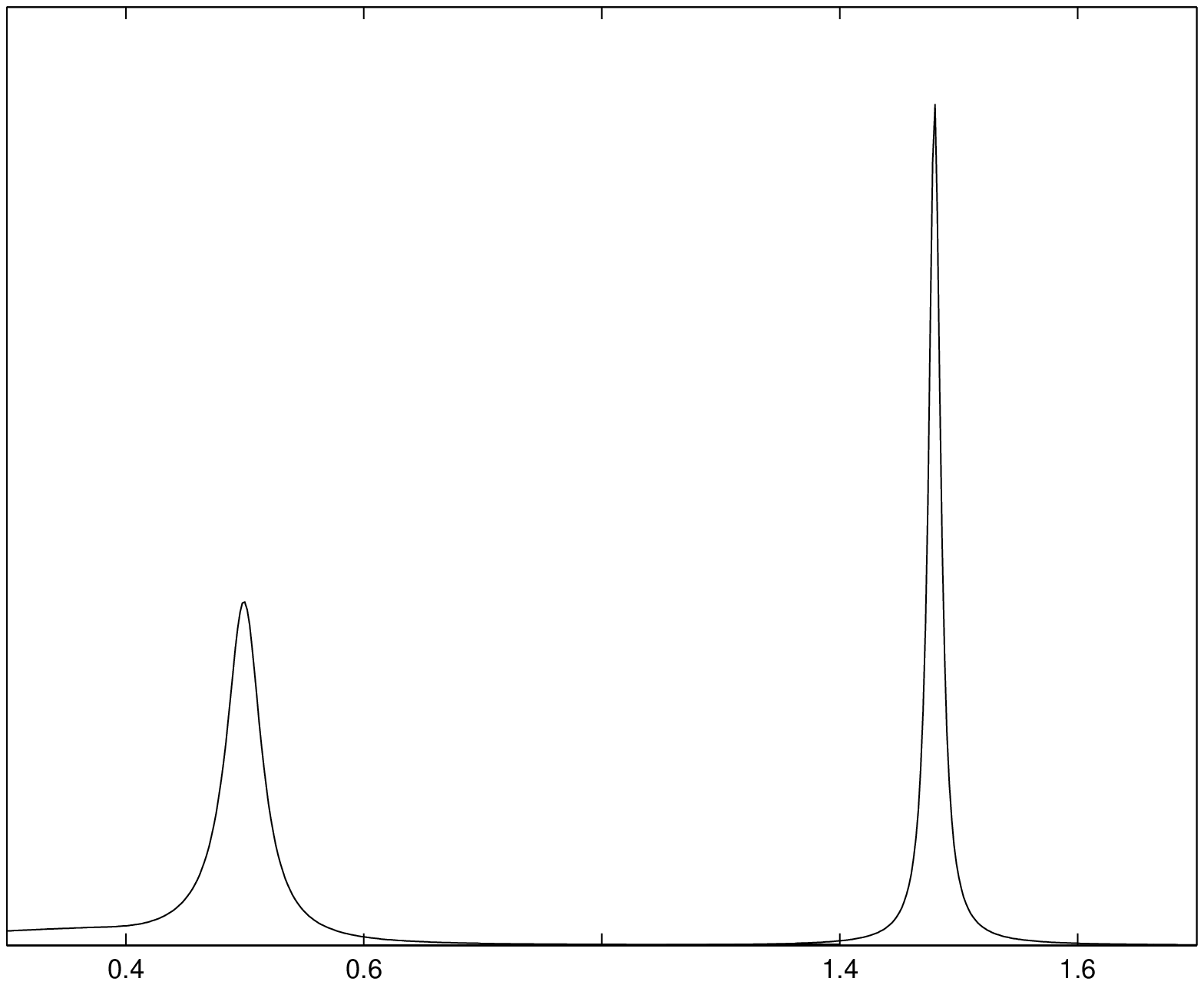}
\begin{picture}(0,-00)
\setlength{\unitlength}{1.000pt}
\put(-130,280){\Large $\omega_{+}$}
\put(-280,120){\Large $\omega_{-}$}
\put(-360,250){\Large $k_{0}=1$}
\put(-360,220){\Large $k=0.4$}
\put(-280,100){\Large $\leftarrow$ }
\put(-260,100){\Large $\times  10$}
\put(-415,100){\rotatebox{90}{intensity}}
\put(-260,-20){frequency transfer, rel. units}
\end{picture}
\end{center}
\caption{ }
\end{figure}

\clearpage
\begin{figure}
\begin{center}
\epsfxsize=140mm
\epsfysize=110mm
\epsfbox{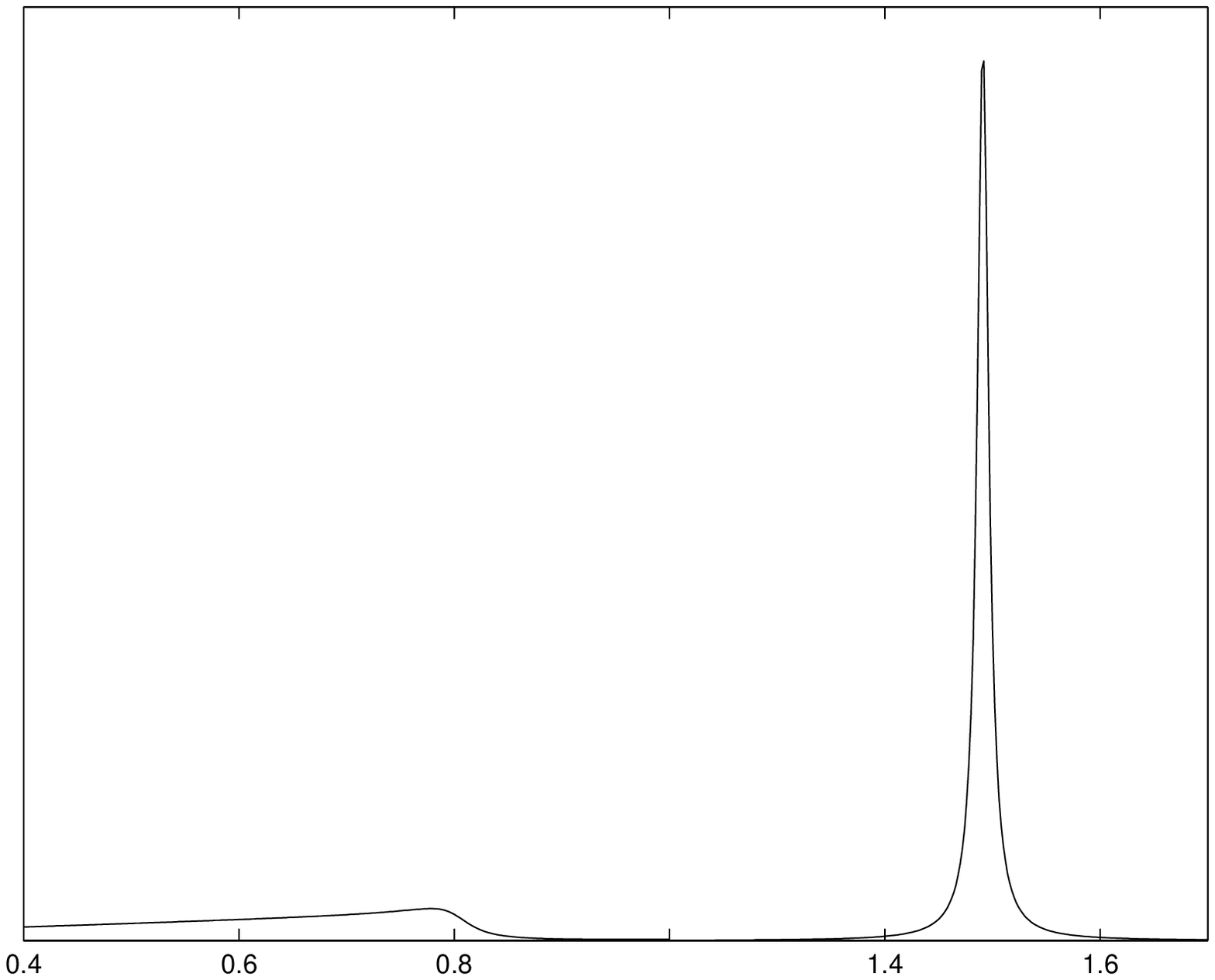}
\begin{picture}(0,-00)
\setlength{\unitlength}{1.000pt}
\put(-130,280){\Large $\omega_{+}$}
\put(-260,100){\Large $\omega_{-}$}
\put(-360,250){\Large $k_{0}=1$}
\put(-360,220){\Large $k= 0.8$}
\put(-260,70){\Large $\downarrow$ }
\put(-240,70){\Large $\times  10$}
\put(-415,100){\rotatebox{90}{intensity}}
\put(-260,-20){frequency transfer, rel. units}
\end{picture}
\end{center}
\caption{ }
\end{figure}

\clearpage
\begin{figure}
\begin{center}
\epsfxsize=140mm
\epsfysize=110mm
\epsfbox{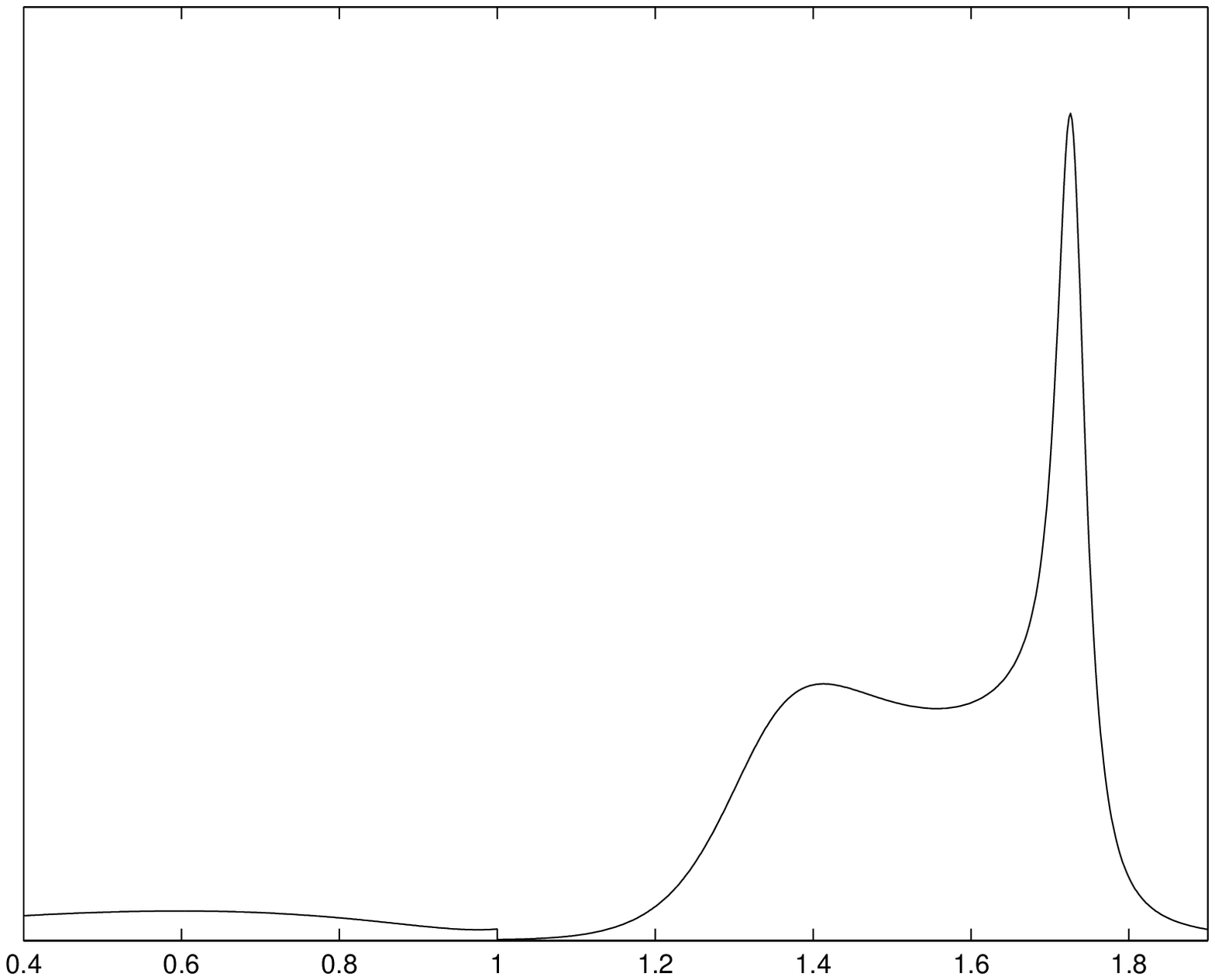}
\begin{picture}(0,-00)
\setlength{\unitlength}{1.000pt}
\put(-100,280){\Large $\omega_{+}$}
\put(-160,120){\Large $\omega_{-}$}
\put(-360,250){\Large $k_{0}=1$}
\put(-360,220){\Large $k= 1.7$}
\put(-350,50){\Large $\downarrow$ }
\put(-330,50){\Large $\times  10$}
\put(-415,100){\rotatebox{90}{intensity}}
\put(-260,-20){frequency transfer, rel. units}
\end{picture}
\end{center}
\caption{ }
\end{figure}


\begin{thebibliography}{30}
\bibitem{Mi} A. B. Migdal,  Zh. Eksp. Teor. Fiz. {\bf 34}, 1438 (1958)
[Sov. Phys. JETP {\bf 7}, 996 (1958)].
\bibitem{BHK} M. Born and Kung Huang, {\it Dynamical Theory of
Crystal Lattices} (Oxford University Press, New York, 1954).
\bibitem{BK} E. G. Brovman and Yu. Kagan, Zh. Eksp. Teor. Fiz. {\bf 52}, 557
(1967) [Sov. Phys. JETP {\bf 25}, 365 (1967)].
\bibitem{AS} A. S. Alexandrov and J. R. Schrieffer, Phys. Rev. B {\bf
56}, 13731 (1997).
\bibitem{AGD}  A. A. Abrikosov, L. P. Gor'kov, I. Ye. Dzyaloshinskii, {\it
Methods of Quantum Field Theory in Statistical Physics}, (Prentice-Hall,
Englewood Cliffs, NJ, 1963).
\bibitem{R} M. Reizer, Phys. Rev. B {\bf 61}, 40 (2000).
\bibitem{GLF}V. L. Gurevich, A. I. Larkin, and Yu. A. Firsov,
  Fiz. Tv. Tela {\bf 4}, 185 (1962).
\bibitem{Kon} V. M. Kontorovich, Usp. Fiz. Nauk {\bf 142}, 265 (1984)
[Sov. Phys. Uspekhi {\bf 27}, 134 (1984)].
%\bibitem{Fa} L. A. Falkovsky, Phys. Rev. B {\bf 66}, 020302-1 (2002).
\bibitem{Fa1} L. A. Falkovsky, Zh. Eksp. Teor. Fiz. {\bf 122}, 411 (2002)
[Sov. Phys. JETP {\bf 95}, No 2(8) (2002)].
\bibitem{AMG} M. d'Adusto, P. K. Mang, P. Giura, A. Shukla,
P. Chigna, A. Mirone, M. Braden, M. Greven, M. Krish, and F. Sette,
cond-mat/0201501 (2002).
\end{thebibliography}
\end{document}